\documentclass[a4paper,12pt]{article}
\usepackage{blindtext}
\usepackage[T1]{fontenc}
\usepackage[utf8]{inputenc}
   \usepackage{authblk}
\usepackage[english]{babel}     
\usepackage{graphicx,wrapfig,lipsum}       
\usepackage[mathscr]{eucal}  
\usepackage{bm}               
\usepackage{amssymb}         
\usepackage{amsmath}      
\usepackage{mathrsfs}      
\usepackage{hyperref}
\usepackage{color} 
\usepackage[normalem]{ulem}
\begin{document}  
\title{\textbf{Study of Gribov Copies in the Yang-Mills ensemble} }

 \author{ \textbf{D.~Fiorentini}\thanks{diego\_fiorentini@id.uff.br}\,\,,
 \textbf{D.~R.~Junior}\thanks{davidjunior@id.uff.br}\,\,,
  \textbf{L.~E.~Oxman}\thanks{leoxman@id.uff.br}\,\,,
 \textbf{G.~M.~Sim\~oes}\thanks{gustavomoreirasimoes@id.uff.br }\,\,,
   \textbf{R.~F.~Sobreiro}\thanks{rodrigo$\_$sobreiro@id.uff.br}\,\,,\\ 
 {\small   \textnormal{  \it
 UFF -- Universidade Federal Fluminense,}}\\
{\small \textnormal{ \it Instituto de F\'{\i}sica, Campus da Praia Vermelha,}}\\
\small \textnormal{ \it Avenida Litor\^anea s/n, 24210-346, Niter\'oi, RJ, Brasil.}\\
} 	
\maketitle
  
\begin{abstract}

Recently, based on a new procedure to quantize the theory in the continuum, it was argued  that Singer's theorem points towards the existence of a Yang-Mills ensemble. In the new approach, the gauge fields are mapped into an auxiliary field space used to separately fix the gauge on different sectors labeled by center vortices. In this work, we study this procedure in more detail. We provide examples of configurations belonging to sectors labeled by center vortices and discuss the existence of nonabelian degrees of freedom. Then, we discuss the importance of the mapping injectivity, and show that this property holds infinitesimally for typical configurations of the vortex-free sector and for the simplest example in the one-vortex sector. Finally, we show that these configurations are free from Gribov copies.

\end{abstract}  
 

\section{Introduction}    

After Singer's theorem \cite{Singer:1978dk}, it became clear that the usual Faddeev-Popov procedure to quantize non-Abelian Yang-Mills theories must be somehow modified in the non-perturbative regime.
Because of a topological obstruction, there is no covariant condition $g(A)=0$ that can globally fix the gauge on the whole configuration space $\{ A_\mu\}$. Hence, when such  condition is imposed, the path integral still contains redundant degrees of freedom (d.o.f) associated with gauge fields obeying $g(A)=0$ and related by nontrivial gauge transformations. Such spurious configurations are typically called Gribov copies. The usual way to deal with this obstruction was implemented in the Landau gauge by V.~N.~Gribov in his seminal work \cite{Gribov:1977wm}, see also Ref.  \cite{Sobreiro:2005ec}. In his proposal,  a path-integral restricted to a subset of $\{ A_\mu \}$  was implemented so as to eliminate infinitesimal copies. As a consequence, the perturbative gauge propagator is destabilized, giving place to one with complex poles, while the ghost propagator is enhanced.   Later on, many other developments were achieved. In the Landau gauge, D.~Zwanziger was able to construct a local and renormalizable action  \cite{Zwanziger:1989mf} which was afterwards refined by the inclusion of dimension two condensates  \cite{Dudal:2008sp,Dudal:2011gd}. Beyond this gauge, it is worth mentioning important progress in the maximal Abelian gauge \cite{Capri:2005tj,Capri:2006cz,Capri:2010an} and the linear covariant gauges \cite{Sobreiro:2005vn,Capri:2015pja,Capri:2016aif}, see also Ref. \cite{Pereira:2013aza}. Finally, we refer to a Becchi-Rouet-Stora-Tyutin (BRST) invariant formulation of the path integral restriction, with a local and renormalizable action, that was implemented as a gauge independent recipe  \cite{Capri:2015ixa,Capri:2016aqq,Capri:2015pfa}.

In Ref.  \cite{Oxman:2015ira},  a different procedure to deal with Singer's obstruction was introduced, by splitting the configuration space into domains $\vartheta_\alpha \subset \{ A_\mu\}$ where local sections are well-defined.  Of course, Singer's theorem does not pose any problem to define regions with a local section having no Gribov copies. The important point is that, in order for these regions to serve as a basis to implement the new proposal, they must form a partition   
\begin{gather} 
\{ A_\mu \} = \cup_{\alpha} \vartheta_\alpha  \makebox[.5in]{,} \vartheta_\alpha \cap \vartheta_\beta = \emptyset ~~ {\rm if}~~ \alpha \neq \beta \;.
\end{gather}
In that case,  we  would have ($S_{\rm YM} = \frac{1}{4g^2} \int d^4x\, F_{\mu \nu}^2$)
 \begin{align}
     Z_{\rm YM}=\sum_{\alpha} Z_{(\alpha)} \makebox[.5in]{,} \langle O \rangle _{\rm YM} =  \sum_\alpha \frac{Z_{(\alpha)}}{Z_{\rm YM}} 
     \langle O \rangle_{(\alpha)}
     \label{ens-cont}
 \end{align}
  \begin{align}
     Z_{(\alpha)} = \int_{\vartheta_\alpha}  [DA_\mu] \, e^{-S_{\rm YM}[A]}
     \makebox[.5in]{,} \langle O \rangle_{(\alpha)} =   \frac{1}{Z_{(\alpha)}} \int_{\vartheta_\alpha} [DA_\mu] \, e^{-S_{\rm YM}[A]} O[A]   \;,
 \end{align} 
 and the usual Fadeev-Popov procedure could be separately implemented on each domain $\vartheta_\alpha$. 
  
  In Ref.  \cite{Oxman:2015ira}, motivated by lattice procedures used to detect center vortices 
  by looking at the lowest eigenfuntions of the adjoint covariant Laplacian \cite{DEFORCRAND2001557,Faber_2001}, the partition of $\{ A_\mu \}$ was generated by correlating $A_\mu$ with the 
 solution $\psi= \psi(A)$  to a set of classical equations of motion
\begin{gather}
\frac{\delta S_{\rm aux}}{\delta\psi_I} =0   \makebox[.5in]{,} \psi_I \in \mathfrak{su}(N) 
 \makebox[.5in]{,} I=1, \dots, N_{\rm f} \;,
\label{eqmot} 
\end{gather}
with appropriate boundary and regularity conditions.  The  $N_{\rm f}$ flavors of auxiliary fields are  adjoint scalars organized in the tuple $ \psi = (\psi_1 , \dots, \psi_{N_{\rm f}})$.  
 Since the auxiliary action $S_{\rm aux} $ is gauge invariant, when an orbit of $A_\mu$ is followed, an orbit in the auxiliary space $\{ \psi \}$ is described, with components
\begin{gather}
\psi_I (A^U) = U \psi_I (A)\, U^{-1}   \makebox[.5in]{,} A_\mu^U = U A_\mu U^{-1} + i U \partial_\mu U^{-1} \;. 
\end{gather}
 Then, a polar decomposition of the tuple $ \psi $ was introduced,
 \begin{gather}
   \psi_1=Sq_1S^{-1},~  \dots ~, \, \psi_{N_{\rm f}}=Sq_{N_{\rm f}}S^{-1} \;,
 \label{polar-dec}
 \end{gather}  
 based on a concept of  ``pure modulus'' condition for a tuple  $ q = (q_1, \dots , 
 q_{N_{\rm f}}) $: 
 \begin{gather} 
  f(q)=0 \makebox[.5in]{,} f \in \mathfrak{su}(N) \;. 
 \label{pure-mod} 
 \end{gather} 
In general, a relation between tuples of the form given in Eq. \eqref{polar-dec} shall be simply denoted by $\psi = q^S$.
When moving along the orbit of $A_\mu$, the modulus $q_I(A)$ stays invariant while the phase describes an orbit $S(A^U) =  US(A)$. Then, the gauge was fixed by requiring $A_\mu$ to be associated with a reference mapping $S_0$ on this orbit. If $S(A)$ is a univocally defined functional, no copies can show up: $S(A) =S_0 $ and $S(A^U)= S_0 $ imply $U \equiv \mathbb{I}$. The reference mapping $S_0$ cannot be globally defined on $\{ A_\mu\}$.  Even for  smooth finite-action configurations $A_\mu$,  $S(A)$ will generally contain defects, which cannot be removed by means of the regular $U$-mappings associated with gauge transformations. Therefore, 
we were led to define classes of equivalent $SU(N)$-mappings, related by left multiplication with a regular $U$. Each class gives rise to  a region $\mathscr{V}(S_0) $ formed by gauge fields such that $S(A) \sim S_0$, where $S_0$ is a class representative. In other words, 
the proposed gauge fixing can only be implemented locally on $\{ A_\mu \}$. This is the reason why it is possible to have no copies in this setting, which is in line with Singer's theorem.

As all the gauge fields belong to some region, and a gauge field $A_\mu$ cannot be in different regions, the above procedure gives a partition of $\{ A_\mu \}$:  $\vartheta_\alpha \to \mathscr{V}(S_0) $. 
The labels correspond to oriented and nonoriented center vortices with nonabelian degrees of  freedom (d.o.f.), where the nonoriented component is generated by monopoles. Then, the YM field averages in Eq. \eqref{ens-cont} involve an ensemble integration over topological defects (sector labels) with a weight $Z_{(S_0)}/Z_{\rm YM} $ that is in principle calculable.  Indeed, the all-orders perturbative renormalizability of the vortex-free sector was shown in Ref. \cite{PhysRevD.101.085007}. The calculation of each sector, 
followed by the ensemble integration, is expected to give rise to the confining behavior in the nonperturbative regime. In this regard, these topological degrees have been established in the lattice as relevant for the confinement of quark probes  in pure $SU(N)$ YM theory  \cite{DelDebbio:1996lih,Langfeld:1997jx,DelDebbio:1998luz,Faber:1997rp,deForcrand:1999our,Ambjorn:1999ym,Engelhardt:1999fd,Engelhardt:1999xw,Bertle:2001xd,Reinhardt:2001kf,Gattnar:2004gx}. Moreover, a measure based on them has recently led to an effective model that captures the asymptotic properties of confinement \cite{oxman4d}.  Of course, this program tends to be prohibitively hard in the continuum. Nonetheless, understanding some of their facets could shed light on how to organize an approximation scheme on each sector. For example, in the calculation of  quadratic fluctuations around a straight thin center-vortex, different self-adjoint extensions are possible \cite{PhysRevD.68.025001}. Which one to use should be determined from first principles and on physical grounds. This  could also provide a guide to compute the different    
sectors in the lattice. The gauge-fixing method is based on many underlying assumptions. In this work, we aim at discussing them at the classical level, paying special attention to sectors that include center vortices.  The purpose is to improve the understanding of the consistency of this procedure.  In our proposal for continuum YM theories, this analysis was still lacking.

In Section 2.  we review the gauge fixing procedure for $SU(N)$, clarifying the necessary assumptions involved. Next, in Sec. 3, we present examples of configurations that belong to sectors labeled by center vortices, and discuss the existence of nonabelian d.o.f.. In Sec. 4 we study the injectivity of the functional $\psi(A)$, and show that it is infinitesimally injective for typical configurations in the vortex-free sector, and for a particular configuration of the one-vortex sector. Finally, in Sec. 5, we establish the absence of Gribov copies for the configurations considered in Sec. 4.

\section{Some general remarks}\label{general}

Let us initially compare the usual and the new gauge-fixing procedures. In the usual case, a  unique  condition
\begin{gather}
   g(A)=0 \makebox[.5in]{} 
\end{gather}
 to pick one point from each orbit in $  \{ A_\mu \}$ is attempted.  Of course, due to Singer's (no-go) theorem \cite{Singer:1978dk}, it is impossible to find a global, continuous, covariant condition such that $g(A^U)=0 \Rightarrow U= \mathbb{I} $. In the new procedure \cite{Oxman:2015ira}, if the solution to Eq. \eqref{eqmot} is unique (after imposing regularity and boundary conditions), in a first step we may associate each field in $\{ A_\mu \} $ with the auxiliary tuple $\psi(A)$  that minimizes the auxiliary action $S_{\rm aux}[A,\psi]$. 
 For this mapping to be useful to fix the gauge, a necessary condition is that different gauge fields of the same orbit are associated with different $\psi(A)$. If $\psi(A)$ is left invariant by nontrivial transformations $\psi_I(A) \to U \psi_I(A) U^{-1}$, $ U \in SU(N)$, then 
no matter what the second step is, the final prcedure will have gauge copies. 
If this is successful, in a second step, given a tuple $\psi(A)$, we would like to fix the gauge by imposing a condition that is satisfied by only one $\psi(A)$ as we move on the orbit of $A_\mu$. 
Again, because of Singer's theorem, it is impossible to find a global and continuous gauge-fixing condition. However, the idea is to implement a different gauge-fixing condition  on each sector of a partition of $\{ A_\mu\}$. As discussed in the introduction, a partition  can be induced  by means of a univocally defined polar decomposition of the tuples $\psi(A)$ in terms of a ``modulus'' tuple $q(A)$ and phase $S(A)$. The sectors $\mathscr{V}(S_0) $ are formed by gauge fields such that  $S(A) \sim S_0$, where $S_0$ is a class representative. 
In summary, for the gauge fixing to be well-defined, we need:

%
\begin{enumerate}
\item Appropriate regularity and boundary conditions on the auxiliary fields so as to have a unique solution $\psi(A)$ to Eq. \eqref{eqmot};  

    \item 
     The auxilary-field content  and the auxiliary action must be such that
$\psi(A)$ is injective on any gauge orbit. This means,
\begin{gather}
\psi(A^U)=\psi(A) \Rightarrow U \in  Z(N)  \;,
\label{inject}
\end{gather} 
where $Z(N)$ is the center group of $SU(N)$;
       \item
      A univocally defined polar decomposition of $\psi(A)$. In this case, besides inducing the partition $\mathscr{V}(S_0) $, a local condition on $\mathscr{V}(S_0) $ with no copies,
\begin{gather}
    f_{S_0}(\psi(A))=0  \makebox[.5in]{\text{and}}  f_{S_0}(\psi(A^U))=0 \Rightarrow U= \mathbb{I} \;, 
    \label{gf1}  
\end{gather} 
would be implemented in terms of the ``pure modulus'' concept in Eq. \eqref{pure-mod} with:
\begin{gather}
    f_{S_0}( \psi)=  f(S_0^{-1} \psi_1 S_0, \dots ,  
 S_0^{-1} \psi_{N_{\rm f}}S_0)  \;,
    \label{gf2}  
\end{gather}  
whose solution is given by $\psi = q^{S_0}$.

\end{enumerate} 

If these requirements are fulfilled, we would have an $S_0$-dependent gauge-fixing condition, without copies on each local sector $\mathscr{V}(S_0) $ of the partition of $\{A_\mu \}$. As in other gauge-fixing procedures, the main idea is not to arrive at a closed expression for the gauge-fixed field $A^{\rm g.f.}_\mu$. This could only be done for some specific cases. In fact, the objective is to properly quantize YM theory. Here, we shall briefly comment about the above requirements,  relating them with the quantization procedure introduced in Ref. \cite{Oxman:2015ira}. A detailed analysis will be developed in the next sections.

Regarding item 1, the natural regularity condition is to consider continuous single-valued  auxiliary fields. In addition, as the gauge fields $A_\mu$ with finite YM action are asymptotically pure gauge,  the natural boundary condition is that $\psi$ is covariantly constant at infinity,
\begin{gather} 
D_\mu \psi \to 0  \makebox[.7in]{\rm when}  |x| \to \infty     \;.
\end{gather} 
 This is consistent with the equations of motion if $\psi(x) \to \bar{\psi} (x) \in \mathcal{M}$ in this limit, where   $\mathcal{M}$ is the vacua manifold of $S_{\rm aux}$. In Ref. \cite{Oxman:2015ira}, starting from the pure YM partition function, 
\begin{gather}
Z_{\rm YM} = \int [DA_\mu]\, e^{-S_{\rm YM}[A]} \;,  
\end{gather}
or the YM correlations, we introduced 
auxiliary fields satisfying Eq. \eqref{eqmot} by means of an identity  
\begin{align}
    1=\int[D\psi]\det\left(\frac{\delta^2 S_{\rm aux}}{\delta\psi_I \delta\psi_J}\right) \delta \left( \frac{\delta S_{\rm aux}}{\delta\psi_I}  \right) \;, \label{firstid}
\end{align}
in the integrand of
the $A_\mu$ path-integration. Given $A_\mu$, to correctly implement this identity,  the argument of the $\delta$-functional must have a unique zero, and the quadratic operator in the determinant must be positive definite. This is nothing but the uniqueness requirement, which is met by the 
regularity and boundary conditions discussed above. In addition, the positivity of the quadratic form is related to solutions $\psi(A)$ with minimum auxiliary action. 

Now, the manifold $\mathcal{M}$ must be nontrivial, that is,  
$S_{\rm aux}$ must be constructed with an appropriate spontaneous symmetry breaking 
(SSB) pattern. If not, $\psi(A)$ could  easily take values close to zero in a spacetime region, and the condition \eqref{inject} in item 2 would be violated.  In other words, we need  $\psi(A)$ to be nontrivial almost everywhere to be able to extract information from it.  Indeed, injectivity will be favored if points $\bar{\psi}$ in $\mathcal{M}$ satisfy \eqref{inject}, which corresponds to require an auxiliary action with an $SU(N) \to Z(N)$ Spontaneous Symmetry Breaking (SSB) pattern. For this to happen, a minimum value of $N_{\rm f}=N$ flavors is needed (see Section \ref{bestaction}). In this case, $\mathcal{M}= SU(N)/Z(N) = {\rm Ad}(SU(N))$, where ${\rm Ad}(\cdot)$ stands for the adjoint representation, and $SU(N)$ acts transitively on this manifold. Then, for a univocally defined polar decomposition (item 3), 
the asymptotic boundary condition would be 
\begin{gather}
\psi(x) \to  \bar{\psi}(x) = u^{\bar{S}}  \makebox[.7in]{\rm when}  |x| \to \infty    \;,
\end{gather} 
where  $u$ is the pure modulus tuple in $\mathcal{M}$ and $\bar{S} = \bar{S}(x) $ is only defined at infinity by
\begin{gather} 
A_\mu \to \bar{S} \partial_\mu \bar{S}^{-1}  \makebox[.7in]{\rm when}  |x| \to \infty    \;.
\end{gather} 
Next, to represent the YM quantities in terms of a partition in the local sectors $\mathscr{V}(S_0) $, we introduced a second identity in the integrand of Eq. \eqref{firstid}
 \begin{align}\label{secondid}
1= \sum_{S_0} 1_{S_0} \makebox[.5in]{,}  1_{S_0} =\int [DU]\, \delta ( f_S (\psi))\det (J(\psi)) \makebox[.5in]{,}   S=US_0\;,
 \end{align}  
where $J(q)$ is the Fadeev-Popov operator associated to the condition \eqref{gf1}. According to item 3, the characteristic function $1_{S_0}$ is nontrivial on fields of the form  
 $\psi = q^S$, $f(q)=0$.  As $\psi$ is single-valued, when we get close to the defects of $S_0$, the fields accompanying Lie algebra components rotated by $S_0$ must tend to zero. 
  When restricted to $\mathscr{V}(S_0) $,  in order for the left-hand side of  $1_{S_0}$ in Eq. \eqref{secondid} to be one, there should be a unique $U$ that solves $f_S(\psi)=0$. This is expected to be addressed by the consideration of the $SU(N) \to Z(N)$ SSB pattern and a good definition of polar decomposition with a univocally defined phase (and modulus).

\label{bestaction}
Let us analyze some possibilities for the auxiliary action
\begin{equation}
    S=\int d^4x \left(\langle D_\mu \psi, D_\mu\psi\rangle+V(\psi) \right) \;,
\end{equation}
 initially focusing on the $SU(2)$ case. The kinetic term is understood as containing a sum over the adjoint scalar fields that form the tuple $\psi$. The Killing product is defined between elements of the Lie Algebra according to
\begin{equation}
    \langle X, Y \rangle = {\rm Tr}({\rm Ad}(X){\rm Ad}(Y))\;.
\end{equation}
 As ${\rm Ad}(SU(2))=SO(3)$, the group action on an adjoint scalar field can be pictured as an orthogonal rotation of a three-component vector. Then, noting that any vector is left invariant by an $SO(2)$ subgroup of rotations, we clearly see that it is not possible to produce $SU(2)\to Z(2)$ SSB with a single scalar field. The situation is different if we consider two adjoint scalar fields
\begin{equation}
    S=\int d^4x \left(\langle D_\mu \psi_1, D_\mu\psi_1\rangle+\langle D_\mu \psi_2, D_\mu\psi_2\rangle+V(\psi_1,\psi_2) \right) \;.
\end{equation}
In this case, if the two vectors are linearly independent, it is clear that there will be no set of continuous transformations that leave them invariant.  However, the potential must be chosen carefully. Asymptotically, the scalar fields should tend to $\mathcal{M}$. It is therefore important that we choose $V$ so as the field components of  the tuples in $\mathcal{M}$ are linearly independent vectors. If we choose 
\begin{equation}
    V(\psi_1,\psi_2)=\lambda_1(\langle \psi_1,\psi_1\rangle -v^2)^2+\lambda_2(\langle \psi_2,\psi_2\rangle -v^2)^2 \;,
\end{equation}
pathological configurations satisfying $\psi_1=\psi_2$, with $\langle \psi_1,\psi_1 \rangle = v$, will belong to $\mathcal{M}$. They are left invariant by rotations with axis $\psi_1$. This can be fixed by adding the term $\langle \psi_1, \psi_2 \rangle^2$ (see Ref. \cite{PhysRevD.18.2932}. Then, $\mathcal{M}$ will consist of two orthogonal vectors which are only invariant by $Z(2) \subset SU(2)$ discrete transformations. Had we added $\langle [\psi_1, \psi_2],[\psi_1, \psi_2]\rangle$ instead, the pathology would persist. In general, among the interesting possibilities is the color-flavor symmetric action containing $N^2-1$ $SU(N)$ adjoint scalar fields $\psi_I \in \mathfrak{su}(N)$, $I = 1, \dots, N^2-1$,
\begin{align}
     S_{\rm aux} = \int d^4x \left (\langle D_\mu \psi_I , D_\mu \psi_I\rangle  +\frac{\mu^2}{2} \langle \psi_I,\psi_I\rangle + \frac{\kappa}{3} f_{IJK}\langle \psi_I, \psi_J\wedge \psi_K\rangle+\frac{\lambda}{4}( \langle \psi_A\wedge \psi_B\rangle )^2\right)\; .\label{choice}
\end{align}
 where we introduced the notation $A\wedge B \equiv -i[A,B]$. As argued in \cite{Oxman:2012ej}, this action admits $SU(N)\to Z(N)$ SSB, and is thus a viable candidate for an auxiliary action.

 \section{Properties of the Yang-Mills sectors}\label{sectors}

In this section, we provide explicit examples of gauge field configurations belonging to nontrivial sectors labeled by center vortices. Then, we show that the procedure allows us to identify more general sectors labeled by nonabelian d.o.f.. These are not related to ambiguities, but are in fact physically inequivalent possibilities located at the same center-vortex guiding centers. 
\subsection{Some sectors labeled by a guiding center}
Let us start with some general remarks about thick center-vortex configurations of the form
 \begin{equation}
 \label{vortex}
     A_\mu = g a(x) \partial_\mu  \chi \beta \cdot T \makebox[.5in]{,}  \beta \cdot T \equiv \beta|_q T_q  \;,
 \end{equation}
where $\chi$ is a multivalued angle when we go around some closed surface $\Omega$ (guiding center), the elements $T_q$ ($q=1, \dots, N-1$) are Cartan generators of $\mathfrak{su}(N)$,  and $a(x)$ is a scalar profile that goes to 1 at infinity. In principle, this profile could be any smooth function. However, regularity conditions must be imposed on $a(x)$ to prevent singularities in $A_\mu$ and the associated $F_{\mu\nu}$. First of all, $a(x)=0$ at $\Omega$, otherwise $A_\mu$ would not be well-defined there. Next, we evaluate
 \begin{eqnarray}
     F_{\mu\nu} = (\partial_\mu a \partial_\nu\chi - \partial_\nu a \partial_\mu\chi) \beta\cdot T + a(x) \left[\partial_\mu,\partial_\nu\right]\chi \beta\cdot T\;.
 \end{eqnarray}
 We have $\left[\partial_\mu,\partial_\nu\right]\chi=0$ everywhere except at $\Omega$, where $a(x)=0$, so that we can disregard this term. When probing the behavior of $a(x)$ at points very close to $\Omega$, we can take $\chi=\varphi$, the angle of polar coordinates, with the $z-t$ plane taken as the tangent plane passing throught the nearest point  $x_0\in \Omega$. Consequently
\begin{equation}
    \frac{1}{4}\langle F_{\mu\nu}, F^{\mu\nu}\rangle = \frac{1}{2}\beta\cdot\beta \left(\partial_\mu a \partial^\mu a\partial_\nu \chi \partial^\nu \chi -(\partial_\mu a \partial^\mu\chi)^2\right) = \frac{1}{2\rho^2}\beta\cdot\beta\left(\partial_\mu a \partial^\mu a-(\hat{\varphi}\cdot\nabla a)^2\right)\;.\label{thindiv}
\end{equation}
 If we expand $a(x) = a^{(1)}(\varphi,z,t)\rho+a^{(2)}(\varphi,z,t)\rho^2+...$, we must impose $a^{(1)}(\varphi,z,t)=0$ or, otherwhise, the action would be infinite due to the divergence of $\langle F_{\mu\nu},F^{\mu\nu}\rangle$ near $x_0$. In other words, on very general grounds, both $a(x)$ as well as its derivative in the local $\rho$ direction should vanish at every point of $\Omega$. In particular, this excludes thin-vortex configurations, as they are associated to an infinite Yang-Mills action density. Thus, within our framework, typical calculations of the partial contribution $Z_{(S_0)}$ to the Yang-Mills ensemble will consist of path-integrals with regularity conditions at the center-vortex guiding centers. This problem is similar to the computation of a Casimir energy, but with conditions imposed on surfaces with higher codimension $d=2$. In ref. \cite{PhysRevD.101.065014}, the dynamical Casimir effect associated to a moving Dirichlet point was discussed for $d=1,2,3$.    The case $d \geq 2$ was found more subtle to deal with, as it is necessary to renormalize the coupling to obtain a finite effective action for the particle. Codimension $d=2$ is particularly interesting as the coupling acquires dependence on an arbitrary mass scale $\mu$. In this case, it was found that the effective action contains a term proportional to $\dot{u}^2$, $u$ being an unitary tangent vector to the particle's trajectory. If we interpret the nontrivial trajectory of the particle as a curved vortex-like object, this term would be associated to  stiffness. It would be interesting to generalize this calculation to gauge theories, This could allow to make contact with the observed properties of  center vortices in the lattice, which display stiffness and tension terms \cite{3dens,oxman4d,oxmanhugo}. It is also worth noting that investigations regarding quantum corrections to the effective action of a thin center-vortex were carried out in Ref. \cite{PhysRevD.68.025001}. In particular, the one-loop correction to the thin vortex energy was shown to vanish for integer fluxes, for a particular choice of self-adjoint extension of the operator accompanying the fluctuations. The physical determination we are proposing here is different, so that the partial contribution of a center-vortex sector should be reexamined.

\subsection{Antisymmetric center vortices with charge $k$} 
 
According to Eq. \eqref{secondid},  if there is a regular transformation $U \in SU(N)$ such that 
\begin{eqnarray}
    [S_0^{-1}U^{-1}(x)\psi_I(x)(A)U(x)S_0,u_I]=0 \;, \forall x\;, \label{conditions0}
\end{eqnarray}
where $u_I=vT_I$, then  $A_\mu \in\mathcal{V}(S_0)$. In particular,  $q_I(x)(A)\equiv S_0^{-1}U^{-1}(x)\psi_I(x)(A)U(x)S_0$ must be single-valued. To proceed, we consider a class of configurations with cylindrical symmetry 
\[
A_\mu^{(k)} = a_{(k)}(\rho) \partial_\mu \varphi \beta^{(k)} \cdot T   \;,
\]
where $\rho$, $\varphi$ are the radial and angular coordinates (indices between parenthesis are not summed).  
The $(N-1)$-tuple $\beta^k$ is proportional to the weight of the k-antisymmetric representation: $\beta^k=2N\sum_{i=1}^k \omega^i$, $k=1,\dots, N-1$, where $\omega^i$ are the weights of the fundamental representation. The profile $a_k(\rho)$ satisfies the regularity conditions $a_k(\rho\to\infty)=1$, $a_k(\rho=0)=0$. The first condition implies that these are in fact thick center-vortices, as they contribute a center element to the $k-$th power for large enough Wilson Loops that link them.   In this case, $\psi(A)$ was obtained in Ref. \cite{oxmangustavo}, and is given by

\begin{eqnarray}
   & \psi^{(k)}_q=h^{(k)}_{qp}V_{(k)}T_pV_{(k)}^{-1} \;, \\&
   \psi^{(k)}_\alpha=\psi^{(k)}_{\bar{\alpha}}=h^{(k)}_\alpha V_{(k)}T_{\alpha}V_{(k)}^{-1}\;, \\&
   V_{(k)}=e^{i\varphi\beta^k\cdot T}\;,
\end{eqnarray}
where the profiles $a,h_{qp}, h_{\alpha}$ satisfy scalar equations, which were solved numerically for $SU(N)$. We now proceed to show that this solution satisfies eq. \eqref{conditions0}, with $U=\mathbb{I}$. Notice that $[\psi^{(k)}_I(x),u_I^{V_{(k)}}(x)]=v\psi^{(k)}_{IJ}(x)V_{(k)}(x)[T_I,T_J]V_{(k)}^{-1}(x)$. As $\psi^{(k)}_{IJ}(x)$ is symmetric, this vanishes for all $x$. Finally, since $h^{(k)}_{\alpha}(0)=0$, for $\alpha\cdot\beta^k\neq 0$, the fields are single valued. Therefore, the  configuration $A_\mu^{(k)}$ contributes to the sector $V_0^{(k)}$ for all values of the parameters consistent with SSB. Additionally, as the profiles $\psi^{(k)}_{IJ}$ do not have other zeros, the configuration cannot contribute to a sector $V_0'=e^{i\chi'\beta'\cdot T}$, if the guiding-center of $\chi'$ is not located in the plane $\rho=0$.
Moreover, for each $k$, the set of roots $\{\alpha_r\}$ that satisfy $\alpha\cdot\beta^k\neq 0$ is different. To see this, consider, without loss of generality, that $k>k'$. Then, for $p\leq k$ we have
 \begin{align}
 &\beta^k \cdot \alpha_{k'p}=-1 \;, \\&
 \beta^{(k')}\cdot \alpha_{(k')p}=0 \;.
 \end{align}
 Since $h_{\alpha_{k'p}}(0)\neq 0$, the solution $\psi_I^{(k')}$ will not contribute to the sector $V_{(k)}$, as it will not be single valued in the $\rho=0$ plane. Similarly, $\psi_I^{(k)}$ will not contribute to $V_{k'}$, as for $q> k'$ 
 \begin{align}
 &\beta^k \cdot \alpha_{k'q}=0 \;, \\&
 \beta^{(k')}\cdot \alpha_{(k')q}=1 \;.
 \end{align}
Therefore, the phases $V_{(k)}$  represent physically inequivalent center-vortex configurations  that are located around the same points in spacetime (they have the same guiding-centers). 

\subsection{Nonabelian degrees of freedom}
  
In nonabelian models with spontaneous symmetry breaking, vortices can have an internal orientational moduli \cite{PhysRevD.71.045010,Hanany_2003,AUZZI2003187}. In our case, although we are dealing with a pure gauge theory, a similar situation occurs when defining the different sectors. As discussed in Ref. \cite{oxman4d}, the multiplication of a general defect $S_0$ by a regular phase $\tilde{U}(x) \in SU(N)$ on the right could yield a physically inequivalent label. For $S_0=e^{i\varphi \beta\cdot T}$, the new configuration is given by
 \begin{gather}
 A_\mu  = a\,  i S \partial_\mu S^{-1} = S \mathcal{A}_\mu S^{-1} + i S \partial_\mu S^{-1}   \makebox[.5in]{,} \mathcal{A}_\mu = (1-a) i S^{-1} \partial_\mu S  \, 
 \end{gather}
\begin{gather}
S= \tilde{U} e^{i \varphi \beta \cdot T} \tilde{U}^{-1}  \;, 
\end{gather} 
while the associated solution can be written   in the form $\psi_A = S \bar{\psi}_A S^{-1}$, where $\tilde{U}$ and $\bar{\psi}_A$ are  single-valued and regular. Using 
 \begin{gather}
S^{-1}  \partial_\mu S = \tilde{U} e^{-i \varphi \beta \cdot T} \tilde{U}^{-1} \partial_\mu \tilde{U} e^{i \varphi \beta \cdot T} \tilde{U}^{-1}
 + i \partial_\mu \varphi  \tilde{U} \beta \cdot T  \tilde{U}^{-1}  
 + \tilde{U} \partial_\mu \tilde{U}^{-1}, 
 \end{gather}
 \begin{eqnarray}
     S^{-1}(D_\mu(A) D^\mu(A) \psi_A)S &=& \square \bar{\psi}_A + 2 \mathcal{A}_\mu \wedge \partial_\mu \bar{\psi}_A  + \partial_\mu \mathcal{A}_\mu \wedge \bar{\psi}_A + \mathcal{A}_\mu \wedge (\mathcal{A}_\mu \wedge \bar{\psi}_A)  \;,
     \label{property}
 \end{eqnarray}
 and  the regularity conditions of $\bar{\psi}_A$ and $a(x)$ to expand $\bar{\psi}_{A} =\bar{\psi}_{A} ^{(0)} + \bar{\psi}_{A} ^{(1)}\rho +...$, $a(x) = a^{(1)}\rho+a^{(2)}\rho^2+...$, we see that the term of order $\rho^{-2}$ in Eq. \eqref{property} is
 \begin{eqnarray}
     \frac{\partial^2 \bar{\psi}_{A} ^{(0)}}{\partial \varphi^2} - 2 \tilde{X} \wedge \frac{\partial \bar{\psi}_{A} ^{(0)}}{\partial \varphi} + \tilde{X}  \wedge (\tilde{X}  \wedge \bar{\psi}_{A} ^{(0)} ) 
     \makebox[.5in]{,} \tilde{X}  = \tilde{U} \beta \cdot T  \tilde{U}^{-1}\;.
 \end{eqnarray}
Since $\bar{\psi}_A$ is single-valued and regular, the zeroth order term $\bar{\psi}_A^{(0)}$ in the $\rho$-expansion cannot depend on $\varphi$. In addition, since the force $\frac{\delta S_{\rm aux}}{\delta \psi_A}$ has no term of order $\rho^{-2}$, 
 at the guiding center it must be verified
  \begin{eqnarray}
 \tilde{X}  \wedge (\tilde{X}  \wedge \bar{\psi}_{A} ^{(0)} ) = 0 \;.
 \end{eqnarray}
 Taking the scalar product with $\bar{\psi}_{A} ^{(0)}$ and using the positivity of the metric,
we get,
  \begin{eqnarray}
 \tilde{X}  \wedge \bar{\psi}_{A} ^{(0)} = 0 \;,
 \end{eqnarray}
 which implies  $\tilde{U}$-dependent regularity conditions on the components of $\bar{\psi}_A$ that do not 
  commute with $\tilde{X} $. In this way, even when considering a $k=1$ fundamental center-vortex with a given guiding-center,  we showed that there are gauge field configurations that belong to a continuum of physically inequivalent sectors of the Yang-Mills theory. These are genuine nonabelian degrees of freedom that must be integrated in the YM ensemble.

\section{Infinitesimal injectivity of  $\psi(A)$}

In this section we shall see that injectivity is related to the positivity of the operator introduced in the identity of Eq. \eqref{firstid}, and to the absence of nontrivial gauge transformations that leave invariant the auxiliary fields. Then, we show that the functional is injective for typical configurations of the vortex-free sector. A particular example in the one-vortex sector is also provided.
\subsection{Conditions for injectivity}
The equations of motion originated from the auxiliary action $\Sigma=\delta S /\delta\psi$ is a functional of $\psi$ and $A_\mu$, $S=S(\psi,A_\mu)$, and it is invariant under an infinitesimal gauge transformation, i.e. $\delta \Sigma=\delta_A \Sigma + \delta_\psi \Sigma=0$, with 
\begin{align}
    &\delta_A \equiv\int\delta A_\mu^a \frac{\delta }{\delta A_\mu^a} \; \makebox[.5in]{,} 
    \delta_\psi \equiv\int\delta \psi_I^a \frac{\delta }{\delta \psi_I^a}\;.
\end{align}
Thus, by acting with a variation $\delta_\psi$ on $S$, we should get the corresponding solution to another gauge field on the same orbit, $A_\mu^U$. Then, we should study if
\begin{equation}
\delta_A \frac{\delta S}{\delta \psi_I^a(x)}=-\delta_\psi \frac{\delta S}{\delta \psi_I^a(x)}=-
\int dy\frac{\delta^2 S}{\delta\psi_I^a(x)\psi_J^b(y)}f^{bmn}\xi^m(y) \psi_J^n(y)=0\;
\end{equation}
has nontrivial solutions. We may multiply this equation by $f^{am'n'}\xi^{m'}(x) \psi_I^{n'}(x)$ and integrate over $x$ to arrive at 
\begin{align}
    &\int dx \; dy \; \frac{\delta^2 S}{\delta\psi_I^a(x)\psi_J^b(y)} v_I^a(x) v_J^b(y) =0 \;, \label{copies} \makebox[.5in]{,} 
    v_I^a(x) = f^{amn}\xi^m(x) \psi_I^n(x)= (\xi(x) \wedge \psi_I(x))|_a \;.
\end{align}
Since $\psi_I^a$ is a minimum of $S$, all the eigenvalues of $ \frac{\delta^2 S}{\delta\psi_I^a(x)\psi_J^b(y)}$ must be positive, as was already required for the identity in Eq. \eqref{firstid} to be well-defined. Therefore, non-trivial solutions for \eqref{copies} are given by
\begin{equation}
    v_I^a=\delta \psi_I^a=0\; .  \label{cop}
\end{equation}
 We see that the lack of injectivity is associated to the existence of nontrivial gauge transformations that leave $\psi_I$ invariant. By using the definitions $\Psi \equiv \psi_A^B$, $X\equiv \xi^A M^A$, $M^A|_{BC}\equiv if^{ABC}$, we can rewrite condition \eqref{cop} for our choice of auxiliary action (Eq. \eqref{choice}) as
\begin{align}
    \Psi X =0 \;. \label{injectivity}
\end{align}
For non-trivial gauge transformations, the solutions to Eq.  \eqref{injectivity} are related to the existence of zero-modes for $\Psi$. Therefore, we conclude that a lack of infinitesimal injectivity would be associated to configurations that satisfy $\det{\Psi}=0$.

\subsection{Vortex-free sector}
For the vortex-free sector, in the limit of large $v$, we expect that $\Psi=  v\mathbb{I} + \epsilon$, where $\epsilon$ is a small matrix. Defining $b(\epsilon)=\det{(v\mathbb{I}+\epsilon)}$, we must show that $b(\epsilon)\neq 0$ for small $\epsilon$. By expanding it, we may write
\begin{align}
    b(\epsilon)\approx b(0)+\frac{\partial g}{\partial\epsilon^a}{\epsilon^a}\;.
\end{align}
Since $b(0)=\det{v \mathbb{I}}= v^{N^2-1}$ is a finite (and large) value, we may conclude that the only solution to Eq. \eqref{injectivity} in this regime is $X=0$. Hence, on the vortex-free sector, injectivity is ensured.

\subsection{Sectors with center-vortices} 
The argument of the vortex-free sector cannot be extended to sectors labeled by vortices, as $\Psi$ will necessarily be far from the identity near their guiding-centers. We may, however, consider a particular example for $SU(2)$. The simplest case is the sector labeled by an antisymmetric vortex with charge k=1.Then, as $\beta=\sqrt{2}$, we have $S_0=e^{i\varphi\sqrt{2} T_1}$, where $\varphi$ is the angle of cylindrical coordinates. For $SU(2)$, the solution $\psi(A)$, when $A$ is a minimum of the action as well, is known to be \cite{PhysRevD.95.025001}
\begin{align}
    &\psi_1 = h_1(\rho)T_1\;, \nonumber \\
    &\psi_{\alpha_1}=h(\rho) S_0 T_{\alpha_1}S_0^{-1}\;, \nonumber\\
    & \psi_{\bar{\alpha_1}}=h(\rho) S_0 T_{\bar{\alpha_1}}S_0^{-1} \;. \label{particular}
\end{align}
In this case, there is only one root $\alpha_1=\frac{1}{\sqrt{2}}$, satisfying $\alpha_1\cdot\beta=1$, and the following relations hold
\begin{align} 
   & S_0 T_{\alpha_1}S_0^{-1}=\cos(\varphi)T_{\alpha_1}-\sin(\varphi)T_{\bar{\alpha_1}}\;, \nonumber\\&
    S_0 T_{\bar{\alpha_1}}S_0^{-1}=\cos(\varphi)T_{\alpha_1}+\sin(\varphi)T_{\bar{\alpha_1}}\;.
    \end{align}
This implies the following $\Psi$ matrix:
\begin{align}
    \begin{pmatrix}
h_1(\rho) & 0 & 0\\
0 & h(\rho) \cos(\varphi) & -h(\rho) \sin(\varphi) \\
0 & h(\rho) \cos(\varphi) & h(\rho) \sin(\varphi)
\end{pmatrix}\;.
\end{align} 
Now, the condition \eqref{injectivity} implies
\begin{align}
  \begin{pmatrix}
0 & h_1(\rho)\xi_3 & h_1(\rho)\xi_2\\
-\xi_3 h(\rho)\cos(\varphi)-\xi_2h(\rho)\sin(\varphi) & \xi_1h(\rho)\sin(\varphi) &\xi_1 h(\rho)\cos(\varphi) \\
-\xi_3h(\rho)\cos(\varphi)+\xi_2 h(\rho)\sin(\varphi) & -\xi_1h(\rho)\sin(\varphi) & \xi_1h(\rho)\cos(\varphi)
\end{pmatrix} = 0 \;.
\end{align}
For $\rho\neq 0$, this gives $\xi_1=\xi_2=\xi_3=0$. The only problematic region is the plane $\rho=0$, which is a region of null measure in $R^4$. The gauge transformations that would leave $\Psi$ invariant, thus leading to the lack of injectivity, should be different from the identity only in this plane. Such transformations are not continuous, so they can be disregarded. The functional $\psi(A)$ is therefore infinitesimally injective in the one-vortex sector for this particular example. 

\section{A polar decomposition without infinitesimal copies} \label{copies}

As discussed in section \ref{general}, the injectivity of $\psi(A)$ does not guarantee that the   
gauge-fixing is free from copies. We still need to show that, for all sectors $S_0$,
 \begin{equation}
      f_{S_0}(\psi(A))= f_{S_0}(\psi(A^U)) = 0 \rightarrow U=\mathbb{I} \;. \label{gaugecond}
 \end{equation}
 We shall see that this condition is related to the absence of zero modes of the operator introduced in the identity of Eq. \eqref{secondid}. For instance, to analyze Eq. \eqref{gaugecond} in the vortex-free sector,  we must show that if
\begin{equation} 
    (q_I\wedge T_I)|_\gamma = f^{aI\gamma}q_I^a =0\;, \label{cop1}
\end{equation}
then there is no gauge transformation with nontrivial parameters $\xi^a$, such that
\begin{align}
    f^{aI\gamma}f^{anm}q_I^n \xi^m=0\; .\label{cop2}
\end{align}
Of course these are just necessary conditions that a problematic tuple should satisfy, as $q_I$ should also minimize the auxiliary action (\eqref{choice}). These algebraic conditions \eqref{cop1},\eqref{cop2} can also be written by using the generators in the adjoint representation:
\begin{align}
    {\rm Ad}(T_A)|_{BC}\equiv M_A|_{BC}=i f^{ABC}\; ,
\end{align}
and of the matrix 
\begin{align}
    Q|_{Ia}=q_I^a \;.
\end{align}
Then, equations \eqref{cop1} and \eqref{cop2} become, respectively,
\begin{eqnarray}
    Tr(M_b Q)=0 \;, \label{puremod} \\
    Tr(M_\gamma M_b Q)\xi^\gamma=0 \;.\label{copiesmod} 
\end{eqnarray}
We may write these conditions as
\begin{align}
   & J^{AB}\xi^B=0 \nonumber \;\makebox[.5in]{,}J^{AB}\equiv Tr(M^A M^B Q)\;, \label{cops}
\end{align}
and conclude that copies are associated with configurations having $\det{J}=0$. In fact, in Ref. \cite{Oxman:2015ira}, the operator $J$ is introduced in the Yang-Mills partition function by means of the Fadeev-Popov procedure (see eq. \eqref{secondid}). It is therefore expected that copies are related to zeros of this determinant.

Let us start by analyzing the above equations for $SU(2)$. In this case, $f^{ABC}=\frac{\epsilon^{ABC}}{\sqrt{2}}$, and the matrices $M$ and $X$ thus read
\begin{align}
&M_1= \begin{pmatrix}
0 & 0 & 0\\
0 & 0 & \frac{i}{\sqrt{2}} \\
0 & -\frac{i}{\sqrt{2}} & 0 
\end{pmatrix}\;\makebox[.5in]{,}  
M_2= \begin{pmatrix}
0 & 0 & -\frac{i}{\sqrt{2}} \\
0 & 0 & 0 \\
\frac{i}{\sqrt{2}} & 0 & 0 
\end{pmatrix} \;\makebox[.5in]{,}
M_3= \begin{pmatrix}
0 & \frac{i}{\sqrt{2}}  & 0 \\
-\frac{i}{\sqrt{2}} & 0 &0\\
0 & 0 & 0 
\end{pmatrix}  \;,\\
&X=\xi^A M^A=\begin{pmatrix}
0 & \frac{i}{\sqrt{2}}\xi_3  & -\frac{i}{\sqrt{2}}\xi_2 \\
-\frac{i}{\sqrt{2}}\xi_3 & 0 &\frac{i}{\sqrt{2}}\xi_1\\
\frac{i}{\sqrt{2}}\xi_2  & -\frac{i}{\sqrt{2}}\xi_1 & 0 
\end{pmatrix} \label{transfadj} \; . 
\end{align}
The pure modulus condition \eqref{puremod} implies that $Q$ is a symmetric matrix, and thus can be parametrized as
\begin{align}
    Q = \begin{pmatrix}
Q_{11} & Q_{12}   & Q_{13}  \\
Q_{12}  & Q_{22}  &Q_{23} \\
Q_{13}  & Q_{23}  & Q_{33}  
\end{pmatrix} \;.
\end{align}
The equation for copies \eqref{copiesmod} then reads
\begin{align}
    &J^{ab}\xi^b =0 \;\makebox[.5in]{,}  \label{eqcopies0} J = \begin{pmatrix}
Q_{22}+Q_{33} & -Q_{12}   & -Q_{13}  \\
-Q_{12}  & Q_{11}+Q_{33}  &-Q_{23} \\
-Q_{13}  & -Q_{23}  & Q_{11}+Q_{22}  
\end{pmatrix} \;\makebox[.5in]{,}\xi = \begin{pmatrix}
\xi^1 \\ \xi^2 \\ \xi^3
\end{pmatrix}\;. 
\end{align}  
In order for the system \eqref{eqcopies0} to have a nontrivial solution, the determinant of $C$ should be 0 (this is a necessary condition). This yields
\begin{align}
    \det{J} =& (Q_{22}+Q_{33})(Q_{11}+Q_{33})(Q_{11}+Q_{22})-2Q_{12}Q_{23}Q_{13}-Q_{12}^2(Q_{11}+Q_{22})-Q_{23}^2(Q_{22}+Q_{33})\nonumber\\&-Q_{13}^2(Q_{11}+Q_{33})=0\;. \label{conditionforcopies}
\end{align}
 \subsection{\bf Study of copies in the vortex-free sector}

In the vortex-free sector, for the general group $SU(N)$, the gauge-fixed functional $q_I(A)$ satisfies
\begin{align}
    q_I(A)\wedge u_I = 0 \; , \\
    q_I(A) \to vT_I\; , x \to \infty \;. \label{goodbc}
\end{align}
If there is a copy, then there exists a gauge transformation $U(x)$ such that
\begin{align}
      q_I^U(A)\wedge u_I = 0 \; , \\ \label{eqcopies}
      U(x) \to \mathbb{I}, x \to \infty \;.
\end{align}
For infinitesimal transformations, equation \eqref{eqcopies} reads 
\begin{align}
     f^{aI\gamma}f^{anm}q_I^n\xi^m=0 \;. \label{infinitesimal}
\end{align}
 In the vortex-free sector, the boundary condition of Eq. \eqref{goodbc} will imply (on the limit of large v) that the fields Q are close to $v\, \mathbb{I}$  everywhere, i.e. $q_I^a = \delta_I^a+\epsilon^a_I$. Eq. \eqref{infinitesimal} thus becomes
\begin{align}
    \xi^\gamma + f^{aI\gamma}f^{anm}\xi^n\epsilon^m_I=0 \;, \\
    \xi^m(\delta^{m\gamma}+f^{aI\gamma}f^{anm}\epsilon^n_I)=0\;. \label{eqns}
\end{align}
This yields a system of $N^2-1$ linear equations in the variables $\xi^a$, with coefficients that will depend on $\epsilon_I^a$, i.e. 
\begin{align}
    M(\epsilon)\xi = 0 \; ,
\end{align}
where $M$    is the matrix of coefficients. For this system to have a nontrivial solution, a necessary condition is 
\begin{align}
k(\epsilon)\equiv \det{M(\epsilon)}=0 \;.
\end{align}
Since $k(\epsilon)$ is polynomial on the infinitesimal parameters $\epsilon_I^a$, we may approximate:
\begin{align}
   k(\epsilon)\approx k(0) + \frac{\partial k(\epsilon)}{\partial \epsilon_I^a}\epsilon_I^a \;.
\end{align}
As $M(0)$ is simply the $(N^2-1) \times (N^2-1)$ identity matrix, we have $k(0)=1$, a finite value. Therefore, , in the large $v$-limit, there are no Gribov copies for the dominant configurations in the vortex-free sector.

\subsection{\bf Study of copies in a general sector}

In a general sector labeled by a defect $S_0$, the functional $\psi_I(A)$ satisfies
\begin{align}
    \frac{\delta S_{\rm aux}}{\delta\psi_I}=0 \;.
\end{align}
For a general $A$ in this sector, $\psi$ will be of the form $\psi_I=US_0q_IS_0^{-1}U^{-1}$, with $U$ regular. The gauge-fixed $A_\mu$ will be associated to $\zeta_I\equiv S_0q_IS_0^{-1}$, and should satisfy
\begin{align}
    \zeta_I(A)\wedge \eta_I= 0 \; , \\
    \eta_I\equiv vS_0T_IS_0^{-1} \;, \\
    \zeta_I(A) \to vS_0T_IS_0^{-1}\; , x \to \infty \;.
\end{align}
If there is a copy, then there exists a gauge transformation $U(x)$ such that
\begin{align}
      \zeta_I^U(A)\wedge \eta_I =(US_0q_IS_0^{-1}U^{-1})\wedge S_0T_IS_0^{-1}=0 \; , \label{s0} \\
      U(x) \to \mathbb{I}, x \to \infty \;.
\end{align}
We may write condition \eqref{s0} in terms of $q_I$:
\begin{align}
    (S_0^{-1}US_0q_I(S_0^{-1}US_0)^{-1})\wedge u_I = 0 \;.
\end{align}
In terms of the matrix $Q$ defined in the previous section, this is
\begin{align}
    R(S_0^{-1}US_0) Q = Q' \;,
\end{align}
with $Q,Q'$ being pure modulus matrices. By defining $\tilde{U}\equiv S_0^{-1}US_0$, we arrive at the conditions that problematic matrices $Q$ should satisfy:
\begin{align}
    R(\tilde{U}) Q = Q' \;, \\
    \tilde{U}(x) \to \mathbb{I}\;, x \to \infty \;.
\end{align}
An important fact that follows from the definition of $\tilde{U}$ is that if $U$ is infinitesimal, so is $\tilde{U}$. This is so because $S_0 \in SU(N)$, so that it preserves the norm of the vector $\xi$. Specifically,
\begin{eqnarray}
    U&=&\mathbb{I}+\xi^A T^A \rightarrow \tilde{U} = \mathbb{I}+(\xi')^A T^A \notag \;,\\
    \xi'&=&R(S_0) \xi \;.
\end{eqnarray}

The equation for infinitesimal copies is therefore the same in all sectors. However, in a general sector there is no reason to believe that $q_I$ will be close to $vT_I$ everywhere, since some of its components must go to zero at the guiding centers of the vortices. Gauge transformations with parameters that are non-zero only in these regions surrounding the guiding-centers of the vortices could, in principle, yield copies. However, as $v$ grows, these regions become increasingly small.

An example of configuration that could yield copies is when $A_\mu=a(\rho)\partial_\mu\varphi\beta\cdot T$, belonging to the sector labeled by a vortex along the z axis. As discussed in \eqref{particular}, for $SU(2)$, the solution for $\psi(A)$ is known. It is of the form
\begin{equation}
    \psi_I=h_{IJ}S_0T_I S_0^{-1} \;.
\end{equation}
This implies
\begin{equation}
    q_I=h_{IJ}T_J\;.
\end{equation}
The associated $Q-$matrix is symmetric, as required by the gauge fixing. For this to admit infinitesimal copies, eq \eqref{conditionforcopies} should be satisfied at some finite region. The necessary condition for the existence of copies is (eq. \eqref{conditionforcopies})
\begin{equation}
    2h(h_1+h)^2=0 \;.
\end{equation}
Since the profiles $h_1(\rho)$ and $h(\rho)$ are positive for all $\rho>0$ (see Ref. \cite{PhysRevD.95.025001}), it is easy to see that this condition is only satisfied at $\rho=0$, which is a region in $R^4$ of null measure. The transformations that lead to copies are not continuous, as they should be nontrivial only in this plane. Then, they should not be considered as associated to gauge transformations. This configuration, therefore, does not admit Gribov copies.

\section{Conclusions}

In this work, we studied the consistency of a recently proposed procedure to fix the gauge on different sectors of the gauge-field configuration space $\{ A_\mu \}$. Unlike the usual procedure, based on a unique gauge-fixing condition and a restriction to the first Gribov region (to avoid infinitesimal copies), our proposal is based on the consideration of different local conditions on the infinitely many sectors of a partition of $\{ A_\mu \}$. These sectors are labelled by oriented and nonoriented center vortices, and the Yang-Mills path-integral measure includes a sum over partial contributions. Our procedure is suited to detect the microscopic features of center vortices in the continuum, which in global gauge-fixing conditions, like the Landau gauge, are effectively seen signaling the breaking of the perturbative regime at the Gribov horizon  \cite{maas,MAAS2007566c}. Each  partial contribution can be associated to a problem  written in a form  closer to the usual one. 
Here, along the way,  we clarified the relevance of the regularity conditions to solve the auxiliary field equations and provide a physical determination of center-vortex sectors. In principle, this is different from considering a thin or thick center-vortex background plus quantum fluctuations. Instead, it is based on path integrating over gauge and auxiliary fields with given singular phases and regularity conditions. 
 We provided explicit examples of thick center-vortex configurations belonging to nontrivial sectors. We also discussed the existence of nonabelian degrees of freedom, which are related to physically inequivalent labels with the same guiding centers. Finally, we showed the absence of Gribov copies for typical configurations of the vortex-free sector and for the simplest example in the sector labelled by a center vortex.  This points 
to the idea that a possibility to deal with the Singer's obstruction to a global 
gauge-fixing is to approach Yang-Mills theories as an ensemble of center-vortex degrees.

 In a future work, it would be interesting to   establish the absence of copies in all sectors and for more general values of the gauge-fixing parameters. This result, together with the all-orders perturbative renormalizability of these sectors, are important steps towards the establishment of the Yang-Mills ensemble in the continuum.

\section*{Acknowledgments}

 The Conselho Nacional de Desenvolvimento Cient\'{\i}fico e Tecnol\'{o}gico (CNPq), the Coordena\c c\~ao de Aperfei\c coamento de Pessoal de N\'{\i}vel Superior (CAPES), and the Funda\c c\~{a}o de Amparo \`{a} Pesquisa do Estado do Rio de Janeiro (FAPERJ) are acknowledged for their financial support.

\end{document}